\documentclass[]{article}

\usepackage{amsmath,amsfonts,amssymb}
\usepackage{graphicx}
\usepackage[colorlinks=true, allcolors=blue]{hyperref}
\usepackage{listings}
\usepackage{subcaption}
\usepackage{geometry}
\renewenvironment{abstract}
{\noindent\textbf{Abstract.}\space}
{}
\title{Dilated Symmetric Difference for Binary Image Comparison}
\author{Sharon Urieli}
\date{}
\begin{document} 
\maketitle

\begin{abstract}
\noindent
The comparison of two binary images is formulated in terms of mathematical morphology.  A new operator, the dilated symmetric difference, is introduced.  It is shown that the dilated symmetric difference effectively detects differences between binary images, provided that the residual alignment error is within specified bounds.   
\end{abstract}


\section{Introduction}
\label{sec:intro}  
 
Existing image comparison methods fail to differentiate between misalignment due to inaccurate registration and differences due to physical change \cite{Chand2025}.  State-of-the-art detection results are achieved by combining registration and detection using diffusion model features \cite{Chellappa2025}.  Registration and detection are similarly combined in \cite{Mesquita2022, Dufresne2022}.  This technical note presents an approach based on set theory \cite{halmos1960} and mathematical morphology \cite{serra1982} for detecting differences in registered binary images with bounded residual alignment error.

\section{Dilated Symmetric Difference}
\label{sec:method}

\newcommand{\hallowstop}[1]{\overset{\scriptstyle\hspace{3pt}\oplus_{#1}}{\triangle}}
\newcommand{\hallowsplus}{\mathbin{\scalebox{1}{\ensuremath{\triangle}}\mkern-12mu\raisebox{0.8pt}{\scalebox{0.55}{\ensuremath{\oplus}}}\,_r}}

Assume two images to be compared have been binarized, let $A$ and $B$ be the binary masks of the  images.  Assume $A$ and $B$ have been aligned imperfectly, and let $\delta_{\text{align}}$ denote the residual alignment error.  

Let $\overline{X} \triangleq 1-X$ denote the complement of $X$.

The difference mask which contains the elements of $A$ that are not in $B$, and the elements of $B$ that are not in $A$, is defined as the symmetric difference $A \triangle B$ \cite{halmos1960}\footnote{In \cite{halmos1960}, the symmetric difference is defined as $(A \setminus B) \cup (B \setminus A)$, which is equivalent to \eqref{eq:symdiff0}.}:
\begin{equation}
\label{eq:symdiff0}
A \triangle B \triangleq (A \cap \overline{B} ) \cup (B \cap \overline{A} )
\end{equation}
The key contribution of this note is to compensate for the residual alignment error $\delta_{\text{align}}$ by dilation \cite{serra1982}. 

The dilated symmetric difference is denoted\footnotemark: 
\begin{equation}
\label{eq:symdiff}
 A \overset{\scriptstyle\hspace{3pt}\oplus_r}{\triangle} B \triangleq (A \cap \overline{B \oplus D_r}) \cup (B \cap \overline{A \oplus D_r})
\end{equation}%
\footnotetext{An alternative notation $\hallowsplus$ was considered, which resembles J.K. Rowling's symbol for the Deathly Hallows.}\noindent 
where $X \oplus D_r$ denotes binary dilation of $X$ by a disk of radius $r$.

It follows that $A \hallowstop{0} B = A \triangle B$.

The dilation expands the pattern boundary by $r$ pixels, therefore, the dilation radius $r$ should be chosen as a tradeoff. \begin{tabular}[t]{@{}l@{\hspace{2em}}l@{}}
Large enough to compensate for $\delta_{\text{align}}$: & $r > \delta_{\text{align}}$,\\
but small enough that: & $d(\text{pattern boundary, detected region}) > r$,\\
and regions enclosed by the pattern satisfy: & $\text{(largest dimension)} > 2r$.
\end{tabular}

\begin{figure}[ht]
\vspace{-0.5cm}
  \centering
  \begin{subfigure}[t]{0.115\textwidth}
    \includegraphics[width=\textwidth]{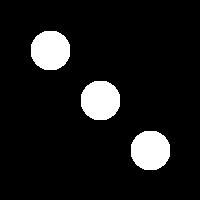}
    \caption{$A$}
  \end{subfigure}
  \begin{subfigure}[t]{0.115\textwidth}
    \includegraphics[width=\textwidth]{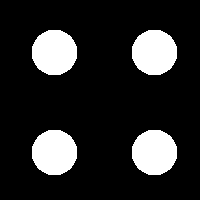}
    \caption{$B$}
  \end{subfigure}
  \begin{subfigure}[t]{0.115\textwidth}
    \includegraphics[width=\textwidth]{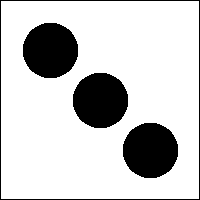}
    \caption{$\overline{A \oplus D_8}$}
  \end{subfigure}
  \begin{subfigure}[t]{0.115\textwidth}
    \includegraphics[width=\textwidth]{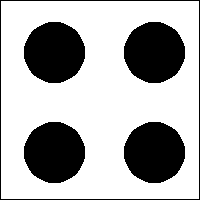}
    \caption{$\overline{B \oplus D_8}$}
  \end{subfigure}
  \begin{subfigure}[t]{0.115\textwidth}
    \includegraphics[width=\textwidth]{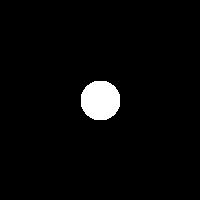}
    \caption{$A \cap \overline{B \oplus D_8}$}
  \end{subfigure}
  \begin{subfigure}[t]{0.115\textwidth}
    \includegraphics[width=\textwidth]{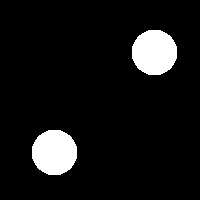}
    \caption{$B \cap \overline{A \oplus D_8}$}
  \end{subfigure}
  \begin{subfigure}[t]{0.115\textwidth}
    \includegraphics[width=\textwidth]{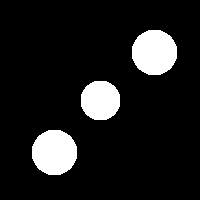}
    \caption{$A \overset{\scriptstyle\hspace{3pt}\oplus_8}{\triangle} B$}
  \end{subfigure}
  \begin{subfigure}[t]{0.115\textwidth}
    \includegraphics[width=\textwidth]{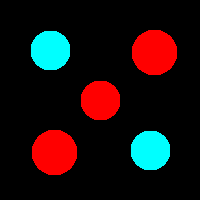}
    \caption{$r=8$}
  \end{subfigure}

 \caption{Die face 3 (die-3) and die face 4 (die-4) share two disk centers.  Alignment errors are introduced by dilating die-4 by $D_3$ and shifting by $(dx,dy)=(4,2)$, relative to die-3, resulting in $\delta_{\text{align}}=3 + \sqrt{4^2 + 2^2} \approx 7.5$.  (a-g) Illustration of the operations of Eq.\eqref{eq:symdiff}. (h) Mask overlay coloring: $B \setminus (A \hallowstop{r} B)$ is shown in cyan, $A \hallowstop{r} B$ is shown in red.  }
  \label{fig:dice}
\end{figure}

\begin{figure}[ht]
  \centering
  \begin{subfigure}[t]{0.24\textwidth}
    \includegraphics[width=\textwidth]{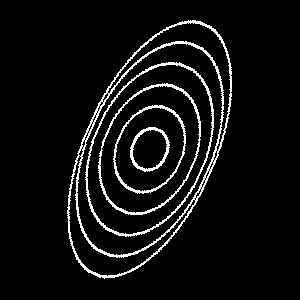}
    \caption{$A$}
  \end{subfigure}
  \begin{subfigure}[t]{0.24\textwidth}
    \includegraphics[width=\textwidth]{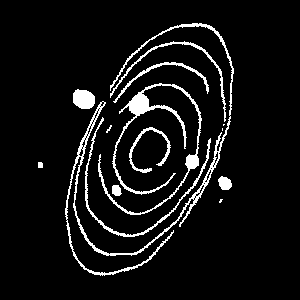}
    \caption{$B$}
  \end{subfigure}
  \begin{subfigure}[t]{0.24\textwidth}
    \includegraphics[width=\textwidth]{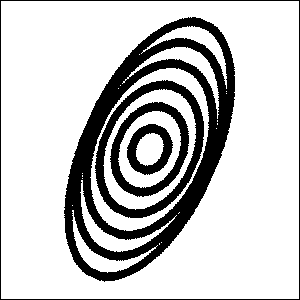}
    \caption{$\overline{A \oplus D_3}$}
  \end{subfigure}
  \begin{subfigure}[t]{0.24\textwidth}
    \includegraphics[width=\textwidth]{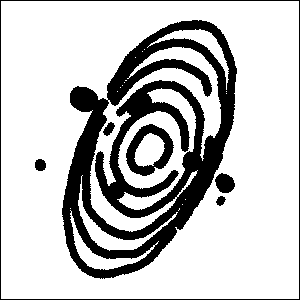}
    \caption{$\overline{B \oplus D_3}$}
  \end{subfigure}
  
  \begin{subfigure}[t]{0.24\textwidth}
    \includegraphics[width=\textwidth]{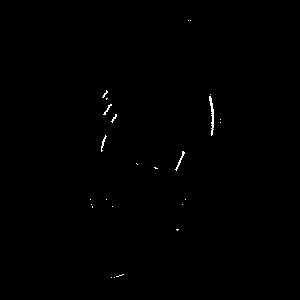}
    \caption{$A \cap \overline{B \oplus D_3}$}
  \end{subfigure}
  \begin{subfigure}[t]{0.24\textwidth}
    \includegraphics[width=\textwidth]{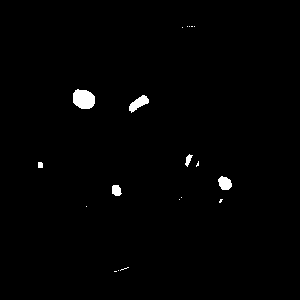}
    \caption{$B \cap \overline{A \oplus D_3}$}
  \end{subfigure}
  \begin{subfigure}[t]{0.24\textwidth}
    \includegraphics[width=\textwidth]{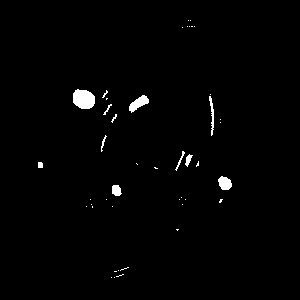}
    \caption{$A \overset{\scriptstyle\hspace{3pt}\oplus_3}{\triangle} B$}
  \end{subfigure}
  \begin{subfigure}[t]{0.24\textwidth}
    \includegraphics[width=\textwidth]{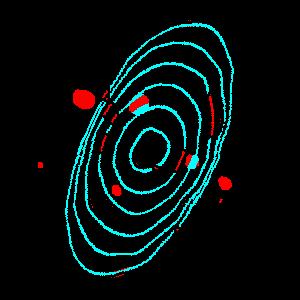}
    \caption{$r=3$}
  \end{subfigure}
  
  \caption{$B$ is misaligned with respect to $A$ by a clockwise $\theta=1^\circ$ rotation and elastic warping with parameters $\sigma=8, \alpha=30$. (a-g) Illustration of the operations of Eq.\eqref{eq:symdiff}. (h) Coloring as in Fig.~\ref{fig:dice}.}
  \label{fig:ellipses}
\end{figure}

Figs.~\ref{fig:dice} and~\ref{fig:ellipses} illustrate the operations of Eq.~\eqref{eq:symdiff} using synthetic examples.

\section{Dilation Radius Analysis}
\label{sec:analysis}

The tradeoff and limitations on the dilation radius $r$ are analyzed using Intersection over Union (IoU\footnote{$\text{IoU}(X,Y)=\frac{|X \cap Y|}{|X \cup Y|}$.}) between $A \hallowstop{r} B$ and a reference mask.
The reference masks are obtained by aligning $A$ with $B$ using the same transformations used to misalign $B$ with respect to $A$ (dilation, shift, rotation and elastic warping \cite{simard2003}).  The exact methods for obtaining the reference masks are given in the figure captions.  

\begin{figure}[ht]
\vspace{-0.5cm}
  \centering
  \begin{subfigure}[t]{0.115\textwidth}
    \includegraphics[width=\textwidth]{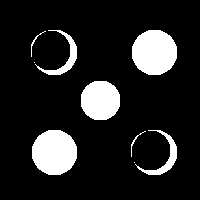}
    \caption{$A \overset{\scriptstyle\hspace{3pt}\oplus_0}{\triangle} B$}
  \end{subfigure}
  \begin{subfigure}[t]{0.115\textwidth}
    \includegraphics[width=\textwidth]{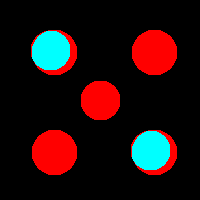}
    \caption{$r=0$}
  \end{subfigure}
  \begin{subfigure}[t]{0.115\textwidth}
    \includegraphics[width=\textwidth]{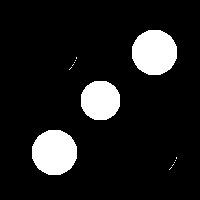}
    \caption{$A \overset{\scriptstyle\hspace{3pt}\oplus_7}{\triangle} B$}
  \end{subfigure}
  \begin{subfigure}[t]{0.115\textwidth}
    \includegraphics[width=\textwidth]{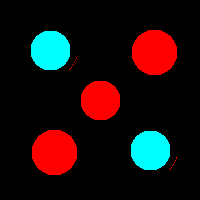}
    \caption{$r=7$}
  \end{subfigure}
  \begin{subfigure}[t]{0.115\textwidth}
    \includegraphics[width=\textwidth]{dice_new_D8_M.png}
    \caption{$A \overset{\scriptstyle\hspace{3pt}\oplus_8}{\triangle} B$}
  \end{subfigure}
  \begin{subfigure}[t]{0.115\textwidth}
    \includegraphics[width=\textwidth]{dice_new_D8_M_overlay.png}
    \caption{$r=8$}
  \end{subfigure}
  \begin{subfigure}[t]{0.115\textwidth}
    \includegraphics[width=\textwidth]{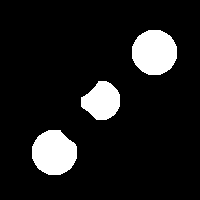}
    \caption{$A \hallowstop{30} B$}
  \end{subfigure}
  \begin{subfigure}[t]{0.115\textwidth}
    \includegraphics[width=\textwidth]{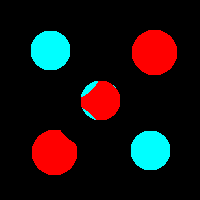}
    \caption{$r=30$}
  \end{subfigure}

  \raisebox{1cm}{%
    \begin{subfigure}[t]{0.115\textwidth}
    \includegraphics[width=\textwidth]{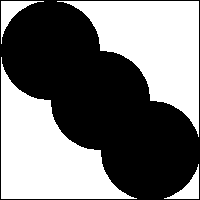}
    \caption{$\overline{A \oplus D_{30}}$}
  \end{subfigure}
  \begin{subfigure}[t]{0.115\textwidth}
    \includegraphics[width=\textwidth]{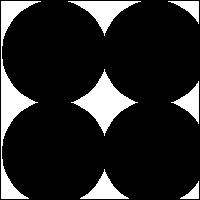}
    \caption{$\overline{B \oplus D_{30}}$}
  \end{subfigure}%
  \hspace{1cm}%
  \begin{subfigure}[t]{0.115\textwidth}
    \includegraphics[width=\textwidth]{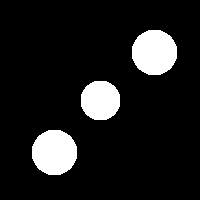}
    \caption{IoU reference mask}
  \end{subfigure}%
  }%
  \hspace{1cm}%
  \begin{subfigure}[t]{0.5\textwidth}
    \includegraphics[width=\textwidth]{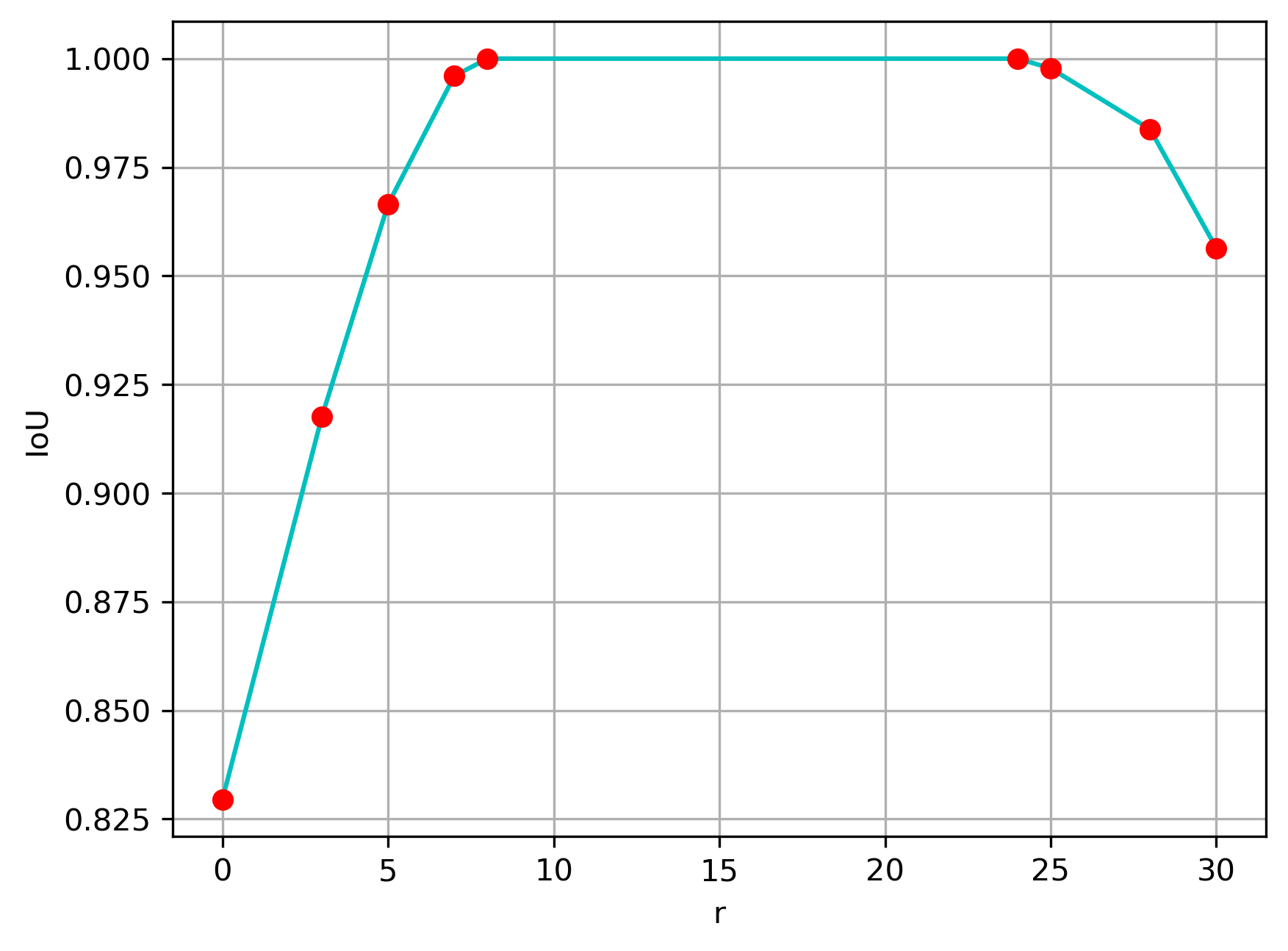}
    \caption{IoU vs. $r$}
  \end{subfigure}

  \caption{(a-b) Symmetric difference without alignment-compensating dilation.  (c-f) The misalignment error disappears when \mbox{$r>\delta_{\text{align}}\approx 7.5$}.  (g-j) $D_{30}$ extends into adjacent disks. (k) The reference mask is obtained by $(\text{die-2}_{dx,dy} \oplus D_3 \triangle \text{die-4}) \cup \text{die-1}$ (die-1 and die-2 not shown). See Fig.~\ref{fig:dice} caption for details.}
  \label{fig:dice_more}
\end{figure}

\begin{figure}[ht]
\vspace{-0.5cm}
  \centering
  \begin{subfigure}[t]{0.23\textwidth}
    \includegraphics[width=\textwidth]{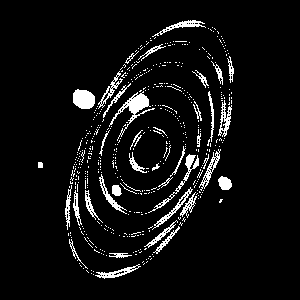}
    \caption{$A \hallowstop{0} B$}
  \end{subfigure}
  \begin{subfigure}[t]{0.23\textwidth}
    \includegraphics[width=\textwidth]{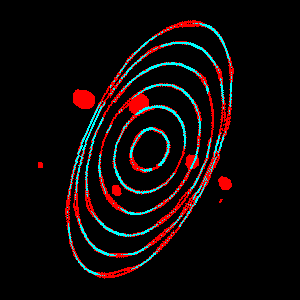}
    \caption{$r=0$}
  \end{subfigure}  
  \begin{subfigure}[t]{0.23\textwidth}
    \includegraphics[width=\textwidth]{fp_R_ellipse_I_elastic_D3_M.png}
    \caption{$A \overset{\scriptstyle\hspace{3pt}\oplus_3}{\triangle} B$}
  \end{subfigure}
  \begin{subfigure}[t]{0.23\textwidth}
    \includegraphics[width=\textwidth]{new_overlay_ellipse_elastic_D3_M_overlay.png}
    \caption{$r=3$}
  \end{subfigure}

  \begin{subfigure}[t]{0.23\textwidth}
    \includegraphics[width=\textwidth]{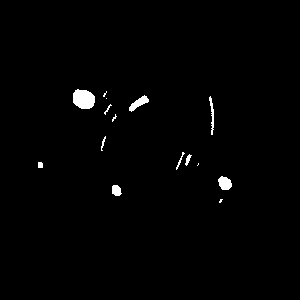}
    \caption{$A \overset{\scriptstyle\hspace{3pt}\oplus_4}{\triangle} B$}
  \end{subfigure}
  \begin{subfigure}[t]{0.23\textwidth}
    \includegraphics[width=\textwidth]{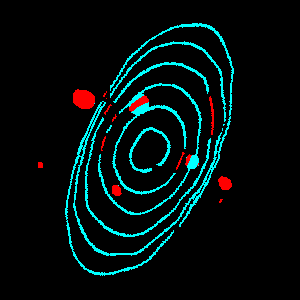}
    \caption{$r=4$}
  \end{subfigure}  
  \begin{subfigure}[t]{0.23\textwidth}
    \includegraphics[width=\textwidth]{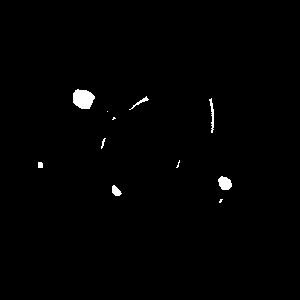}
    \caption{$A \overset{\scriptstyle\hspace{3pt}\oplus_6}{\triangle} B$}
  \end{subfigure}
  \begin{subfigure}[t]{0.23\textwidth}
    \includegraphics[width=\textwidth]{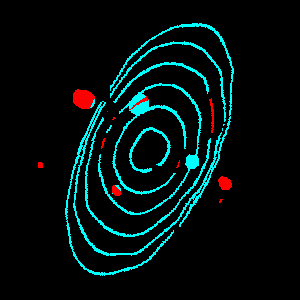}
    \caption{$r=6$}
  \end{subfigure}

\raisebox{0.5cm}{%
  \begin{subfigure}[t]{0.23\textwidth}
    \includegraphics[width=\textwidth]{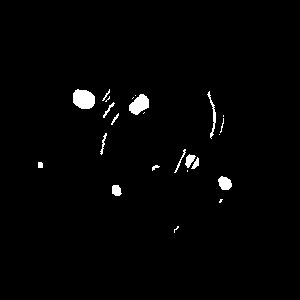}
    \caption{IoU reference mask $A_{\theta \sigma \alpha} \triangle B$ ($\theta \sigma \alpha$ - see Fig.~\ref{fig:ellipses})}
  \end{subfigure}%
}%
\hspace{1cm}%
\begin{subfigure}[t]{0.5\textwidth}
  \includegraphics[width=\textwidth]{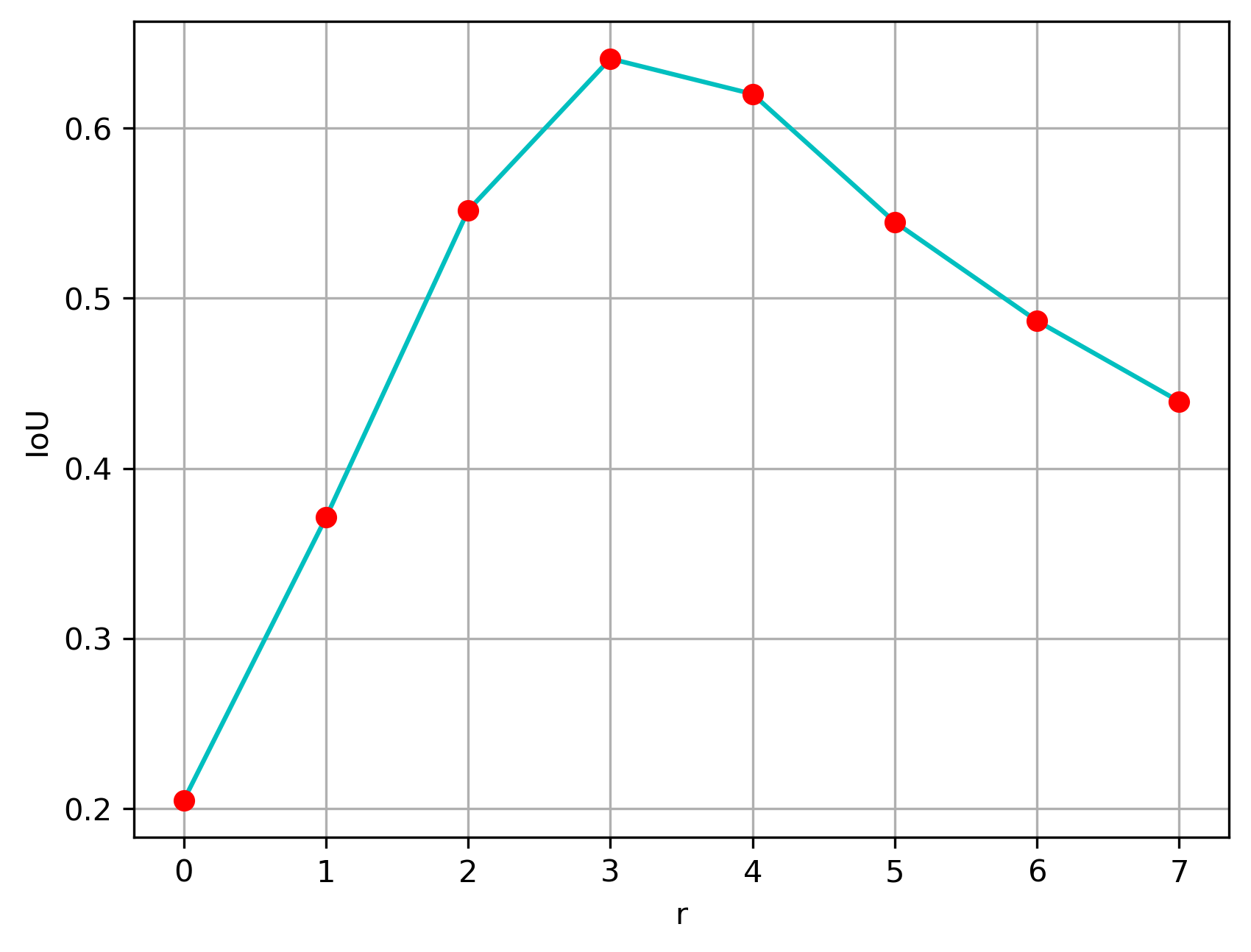}
  \caption{IoU vs. $r$}
\end{subfigure}

  \caption{As $r$ increases, the dilation cuts into the warped disks that are on an ellipse, and the lengths of the detected gaps shrink.  At $r=4$, the misalignment error disappears, but the gaps of length $l\leq 8$ are no longer detected.  When $r=6$, one of the central warped disks is completely engulfed by the dilation.}

  \label{fig:ellipses_more}
\end{figure}

The effects of radius $r$ on the IoU are illustrated in Figs.~\ref{fig:dice_more} and~\ref{fig:ellipses_more}.  In Fig.~\ref{fig:dice_more}, the IoU does not start to fall until $r$ is large enough that $A \oplus D_r$ overlaps with $B$ or $B \oplus D_r$ overlaps with $A$.  The disks are far enough apart that a perfect separation, $\text{IoU}=1.0$, is achieved without tradeoff.  In Fig.~\ref{fig:ellipses_more}, the IoU does not reach 1.0 because the dilations of $A$ and $B$ extend into the detected regions that are close to the pattern.     

\section{Conclusion}
\label{sec: conclusion}

Dilated symmetric difference effectively detects differences between binary images, provided that the residual alignment error is within specified bounds.  A natural extension is to integrate the dilated symmetric difference as a differentiable morphological layer in a neural network \cite{Aouad2022, Shen2019}, enabling the dilation radius, or in the generalized sense the structuring element, to be learned adaptively within a mathematical morphology based pipeline that includes registration. 

\section*{Acknowledgments}
The author thanks A. Urieli for placing this work in current context, T. Eisenberg for questions that led to a more focused presentation and T. Urieli for noting the Deathly Hallows resemblance.

\bibliography{report} 
\bibliographystyle{unsrt}

\end{document}